\documentclass[12pt]{iopart}

\usepackage{iopams}
\begin{document}

\title{Quantum $\widehat{su}(n)_k$ monodromy matrices}

\author{
P Furlan$^{1,2}$ %\footnote[1]{e-mail address: furlan@trieste.infn.it}
and L Hadjiivanov$^{3,2}$ %\footnote[2]{e-mail address: lhadji@inrne.bas.bg}
}

\address{$^1$ Dipartimento di Fisica dell' Universit\`a degli Studi di Trieste, Strada Costiera 11, I-34014 Trieste, Italy}
\address{$^2$ Istituto Nazionale di Fisica Nucleare (INFN), Sezione di Trieste, Trieste, Italy}
\ead{furlan@trieste.infn.it}
\address{$^3$ % Theoretical and Mathematical Physics Division,
Institute for Nuclear Research and Nuclear Energy, Bulgarian Academy of Sciences, Tsarigradsko Chaussee 72, BG-1784 Sofia, Bulgaria}
\ead{lhadji@inrne.bas.bg}

\begin{abstract}
The canonical quantization of the chiral Wess-Zumino-Novikov-Witten (WZNW)
monodromy matrices, both the diagonal and the general one,
requires additional numerical factors that can be attributed to renormalization.

We discuss the field-theoretic and algebraic aspects of this phenomenon for the $SU(n)$
WZNW model and show that these quantum renormalization factors are compatible with the
natural definitions for the determinants of the involved matrices with non-commuting entries.

%This document describes the  preparation of an article using \LaTeXe\ and
%\verb"iopart.cls" (the IOP \LaTeXe\ preprint class file).
%This class file is designed to help
%authors produce preprints in a form suitable for submission to any of the
%journals published by IOP Publishing.
%Authors submitting to any IOP journal, i.e.\
%both single- and double-column ones, should follow the guidelines set out here.
%On acceptance, their TeX code will be converted to
%the appropriate format for the journal concerned.

\end{abstract}

%\pacs{02.10.Yn, 02.20.Uw, 02.40.Gh}
\maketitle

%%%%%%%%%%%%%%%%%%%%%%%%%%%%%%%%%%%%%%%%%%%%%%%%%%%%%%%%%%%%%%%%%%%%

%\hfill 07.11.2011/INFN/ICTP Trieste - FH2-2 %
%\hfill 12.01.2012/INRNE Sofia - FH2JPA %
%\hfill 16.01.2012/INRNE Sofia - FH2JPA %
%\hfill 06.03.2012/INRNE Sofia - FH2JPArevised %

%%%%%%%%%%%%%%%%%%%%%%%% (AMS)LATEX MACROS %%%%%%%%%%%%%%%%%%%%%%%%%

\def\theequation{\thesection.\arabic{equation}}
\def\be{\begin{equation}}
\def\ee{\end{equation}}
\def\ba{\begin{eqnarray}}
\def\ea{\end{eqnarray}}
\def\lb{\label}
\def\nn{\nonumber}

\def\a{\alpha}
\def\b{\beta}
\def\g{\gamma}
\def\d{\delta}
\def\i{\eta}
\def\e{\varepsilon}
\def\l{\lambda}
\def\r{\rho}
\def\s{\sigma}
\def\t{\tau}
\def\o{\omega}
\def\v{\varphi}
\def\x{\xi}

\def\D{\Delta}
\def\G{\Gamma}
\def\O{\Omega}
\def\L{\Lambda}

\def\fU{\mathfrak A}
\def\bo{\mathfrak b}
\def\fp{\mathfrak p}

\def\E{{\cal E}}
\def\Vp{{\cal V}_p}
\def\Hp{{\cal H}_p}

\def\bq{\overline{q}}
\def\bM{\bar M}
\def\bz{\bar z}
\def\bU{\overline{U}_q}
\def\bD{\overline {\cal D}}
\def\tU{{\tilde{U}}_q}

\def\id{\mbox{\em 1\hspace{-3.4pt}I}}
\def\idd{\scriptsize{\em 1 \hspace{-4.2pt}I}}
\newcommand{\ID}[2]{\id^{| #1 {\cal i}}_{\;\;\; {\cal h} #2 |}}

\def\p{\hat p}

\def\Z{\mathbb Z}
\def\R{\mathbb R}
\def\C{\mathbb C}
\def\F{\mathbb F}
\def\H{\mathbb H}

\def\eod{\phantom{a}\hfill \rule{2.5mm}{2.5mm}}

\def\hR{\hat{R}}
\def\Rp{\hat{R}(p)}
\def\subbbc{{\rm C}\kern-3.3pt\hbox{\vrule height4.8pt width0.4pt}\,}
\def\vac{\mid 0 \rangle}

%%%%%%%%%%%%%%%%%%%%%%%%%%%%%%%%%%%%%%%%%%%%%%%%%%%%%%%%%%%%%%%%%%%%%%%%%%%%%%

\section{Introduction}
\label{sec:1}

\setcounter{equation}{0}
\renewcommand\theequation{\thesection.\arabic{equation}}

The WZNW model \cite{WZ, N, W} defined in terms of a simple compact Lie group $G$ (which in our case will be also connected and simply connected)
and a positive integer $k\,,$ the level, is a basic example of a unitary rational conformal field theory (CFT) \cite{DFMS}.
Taken over a cylindric $2D$ space-time (with periodic space), the dynamics of the group valued WZNW field $g(x^0,x^1)\,$
is equivalent to that of a closed string moving on a group manifold \cite{GW}.

Due to the two-sided chiral symmetry of the model, its quantum version can be appropriately formulated in terms of highest
weight/lowest energy representations of two commuting conformal current algebras (see e.g. \cite{FSoT}). The correlation functions can be expressed,
accordingly, as sums of products of (left and right) chiral conformal blocks \cite{BPZ, DF}. The latter are multivalued analytic functions in the
corresponding chiral variables which satisfy the Knizhnik-Zamolodchikov (KZ) equation \cite{KZ, T}. It has been noticed first in \cite{TK, Kohno}
that the ("monodromy") representations of the braid group on the corresponding spaces of KZ solutions are related to the then recently
discovered quantum groups \cite{D}.

The canonical quantization approach to the WZNW model \cite{B, Bl, F1, AS, G, FG1} (see e.g. \cite{FHT1, FHIOPT} and references therein for further
developments) provides an alternative, operator approach to the model. The naive prescription of "replacing the Poisson brackets (PB) by
commutators" is only directly applicable to the commutation relations of the conserved chiral currents.
Those of the chiral components of the (Sugawara type) stress-energy tensor require a well known additive renormalization of the level,
$k \to k+g^\vee =: h\,,$ where $g^\vee$ is the dual Coxeter number of the Lie algebra ${\cal G}$ of $G\,$
and $h$ is the {\em height}. The quadratic PB of
the group valued chiral fields involving classical $r$-matrices are replaced by quantum $R$-matrix exchange relations possessing
the correct quasiclassical asymptotics and appropriate quantum symmetries. To construct the corresponding state space
respecting energy positivity and covariance, one considers vacuum representations of the exchange algebras with a vacuum
vector that would guarantee these properties.

In the canonical framework the chiral splitting requires the introduction of {\em monodromy matrices}
accounting for the quasi-periodicity of the matrix "chiral field operators" (related to the multivaluedness of the
conformal blocks considered as $n$-point functions of their entries). The monodromy matrices are to some extent
a matter of choice and fall essentially in two groups, diagonal ones (belonging to the maximal torus of $G$,
which we will denote by $M_p$) and general, $M\in G\,$ (further restrictions will be discussed in the main text).
It has been shown already in \cite{FHIOPT} that in both cases the monodromy matrices are subject to a quantum
renormalization by specific numerical factors, which are different for $M\,$ and $M_p\,.$ Some algebraic aspects
of the renormalization of $M\,$ (a solution of the {\em reflection equation}) have been discussed in \cite{IP}.

The aim of the present paper is twofold: first, to collect and discuss in detail the {\em field-theoretic}
arguments of the quantum renormalization of the monodromy matrices and second, to provide additional
{\em algebraic} reasons for their presence. To make the paper self-contained, the presentation of the new and possibly
interesting facts and formulas is preceded by a compehensive introduction to the subject (and
supplemented by a rather long list of references) which could be, hopefully, of interest on its own,
containing specific information otherwise scattered in different papers.

\smallskip

We show, in all cases of interest, that the so defined quantum determinants possess the {\em factorization property}
(the determinant of a product is equal to the product of determinants, see Eqs. (\ref{DaDMp}), (\ref{det-mult}) and
(\ref{MMMpm}) below) which is a quite non-trivial fact for matrices with non-commuting entries.

\smallskip

The content of the paper is the following. Section 2 provides a synopsis on the classical WZNW model and its
canonical quantization, with special attention to the case $G=SU(n)\,.$ In the next Section 3 we give a description
of the $SU(n)$ WZNW chiral state space as a collection of representation spaces of the affine algebra
$\widehat{su}(n)_k$ and of the quantum group $U_q$, an $n$-fold cover of $U_q(s\ell(n))$, which plays the role
of internal symmetry (gauge) group. The $U_q$ representation spaces are generated from the vacuum
by the quantum zero modes' matrix $a$ which intertwines between the diagonal monodromy
$M_p\,$ and the general one, $M\,.$ The definition of the quantum determinant ${\det}_q(a)\,,$ introduced in \cite{HIOPT}
(based on ideas of Gurevich et al. concerning Hecke algebras and quantum antisymmetrizers, cf. e.g. the references
in \cite{IP}) is briefly reviewed in Section 4. In Section 5 we provide the field-theoretic reasons for the quantum
renormalization of the monodromy matrices $M_p$ and $M$. In the next sections which are of purely algebraic flavor
we propose natural definitions for the corresponding quantum determinants (the diagonal monodromy is considered
in Section 6, and the general one in Section 7) and prove the factorization property in each of the cases.
In the last Section 8 we prove two important identities following from various $R$-matrix exchange relations.

\section{The classical WZNW model and its canonical quantization}
\label{sec:2}

\setcounter{equation}{0}
\renewcommand\theequation{\thesection.\arabic{equation}}

The general solution \cite{W} of the classical WZNW equations of motion for the periodic $2D$ group-valued field
$g(x^0,x^1) = g(x^0, x^1+2{\pi})\,$ is given by the product of two arbitrary {\em chiral fields},
\be
\lb{LR}
g(x^0, x^1) \, \equiv \, g (x^+ ,x^- ) = g_L (x^+ )\, g_R^{-1} (x^- )\ ,\qquad x^\pm = x^1 \pm x^0
\ee
which are only {\em twisted periodic}:
\be
\lb{cM}
g_L (x^+ +2\pi ) = g_L (x^+)\, M\ ,\quad g_R (x^- +2\pi) = g_R (x^-)\, M\ .
\ee
The way the solution (\ref{LR}) is written down (with the {\em inverse} of $g_R$ \cite{FG1}) makes the
relation between the two chiral sectors quite transparent: the $2D$ symplectic form is a sum of the two chiral ones (sharing the same monodromy)
which only differ in sign, so the same is valid for the corresponding PB.
The chiral symplectic forms are determined up to the addition of a monodromy dependent $2$-form $\rho (M)$ \cite{G}
whose external differential is equal to the WZ term (the canonical $3$-form on $G$)
\be
d\,\rho (M) = \theta (M) := \frac{1}{6}\,{\rm tr}\, ([ M^{-1} d M \stackrel{\wedge}{,} M^{-1} d M ]\wedge  M^{-1} d M )\ ,
\lb{WZM}
\ee
but is arbitrary otherwise. The presence of $\rho (M)$ in both chiral symplectic forms provides for their closability.
However, as $\theta (M)$ is not exact (cf. e.g. \cite{Schw}), such a smooth $2$-form can only be defined locally on $G$.
We will only deal with one chiral WZNW sector (the left one,
denoting henceforth $g_L(x^+)$ by just $g(x)$), paying special attention to the corresponding monodromy matrix.
The entries of $M$ carry dynamical degrees of freedom having, in particular, non-zero PB with $g(x)\,.$
There are, essentially, two options in choosing the submanifold of $G$ to which the monodromy belongs.

The first of these is setting the monodromy matrix to be {\em diagonal}, i.e. to belong to a maximal torus $T\subset G\,.$
The WZ term $\theta (M_p)$ vanishes\footnote{In the {\em complex} case, $\theta (M)$ vanishes exactly when $M^{-1} d M$ takes values in a solvable
subalgebra of the complexification $G_{\subbbc}$ of $G\,$ (cf. e.g. \cite{FS} for the corresponding {\em Cartan criterion}).} so that $\rho(M_p)$ could be just any closed $2$-form.
The corresponding chiral fields are called "Bloch waves". We will use the special notation $u(x)\,,$ for the field, and $M_p\,,$
for its monodromy matrix in this case, so that
\be
\lb{uMp}
u (x +2\pi ) = u (x)\, M_p\ ,\qquad M_p = e^{\frac{2\pi i}{k}\, p}\ ,\qquad i p \in {\frak h}
\ee
where ${\frak h} \subset {\cal G}$ is the Lie algebra of $T$. A convenient parametrization for ${\cal G}=s\ell(n)=A_{n-1}$
is provided by the "barycentric coordinates" $\{ p_i \}_{i=1}^n$ of the weights in the "orthogonal basis" of the
root space\footnote{By this in the $A_{n-1}$ case one understands, as usual, the orthonormal basis $\{ \e_s \}_{s=1}^n$
of an auxiliary $n$-dimensional Euclidean space in which the root space is identified with the hyperplane orthogonal to
$\sum_{s=1}^n \e_s$ and the $A_{n-1}$ simple roots are given by $\a_i = \e_i - \e_{i+1}\,,\ i=1,\dots, n-1\,,$ see e.g. \cite{FS}.}
dual to the diagonal Weyl matrices $e_i\,,\ {\rm tr}\, (e_i\, e_j ) = \d_{ij}\ ,\ i,j = 1,\dots ,n$:
\be
p = \sum_{i=1}^n p_i\, e_i\ ,\quad
{\rm tr}\, p = 0\quad\Leftrightarrow\quad \sum_{i=1}^n p_i = 0\ .
\lb{bary}
\ee
In these coordinates the the fundamental Weyl chamber $C_W$ and the level $k$ positive Weyl alcove $A_W$ can be identified,
respectively, with
\be
C_W = \{ p \, \mid \, p_{i i+1} \ge 0\,,\ i=1,\dots, n-1\}\ ,\quad A_W = \{ p \in C_W \, \mid \, p_{1n}  \le k\ \}
\lb{CAW}
\ee
where $p_{ij} := p_i - p_j\,.$ Redefining $u(x)\,$ by multiplying it from the right by a suitable element of the Weyl group,
the diagonal monodromy $M_p\,$ can be always restricted to $p \in C_W\,.$

The ensuing quadratic PB for the Bloch waves
\be
\fl
\qquad
\{ u_1 (x_1) , u_2 (x_2) \}  = u_1 (x_1) \, u_2 (x_2)\,
( \frac{\pi}{k}\, C_{12} \,  \varepsilon (x_{12} ) - r_{12} (p) )\quad{\rm for}\quad |x_{12}| <2\pi
\lb{PBBW}
\ee
involve the $r$-matrix $r_{12}(p)\in{\cal G}\wedge{\cal G}$ satisfying the classical dynamical Yang-Baxter equation
\cite{GN, BDF, EV},
as well as the polarized Casimir operator $C_{12}$ (characterized by its $ad$-invariance, $[C_{12} , X_1 + X_2 ] = 0\ \,\forall X\in{\cal G}$).
In the specified interval of $x_{12}$ values, the function $\varepsilon (x)$ coincides with the sign function\footnote{The
twisted periodicity (\ref{uMp}) allows to calculate $\{ u_1(x_1) \,,\, u_2(x_2) \}$ outside this region as well; the same
remark applies also to (\ref{PBg}) and (\ref{cM}).}.
Here we prefer the compact tensor product notation to the index one, writing for example
\ba
&&X_1 := X\otimes \id\otimes \id \otimes\dots\ ,\quad X_2 := \id\otimes X\otimes \id\otimes\dots\quad {\rm etc.}\,,\nn\\
&&C_{12} = \sum_{a,b=1}^{\dim{\cal G}} \eta^{ab}(T_a)_1(T_b)_2\qquad (\,\eta^{ab} = {\rm tr}\,(T^a T^b)\ ,\quad {\rm tr} (T_a T^b) = \d^b_a\,)
\qquad\qquad
\lb{XC}
\ea
where $\{ T_a \}_{a=1}^{\dim{\cal G}}$ and $\{ T^b \}_{b=1}^{\dim{\cal G}}$ form dual bases of the Lie algebra ${\cal G}\,.$

Alternatively, one can allow $M$ to take general group values. This means that the chiral symplectic form $\Omega (g , M)$
necessarily contains a non-trivial locally defined $2$-form $\rho (M)$ which determines
the corresponding $r$-matrix $r(M)$ in the PB of two chiral fields (cf. \cite{BFP} for the exact relation).
It appears natural to ask whether one can get rid of the monodromy dependence of the $r$-matrix \cite{F1, G} (for $|x_{12}| <2\pi$),
and the answer \cite{FG1, FHT1} is the following. All possible chiral field PB with a constant $r$-matrix are of the form
\ba
&&\{ g_1(x_1) , g_2(x_2) \} = \frac{\pi}{k}\, g_1(x_1)\, g_2(x_2) \, (C_{12}\, \varepsilon (x_{12}) - r_{12} ) =\nn\\
&&= - \frac{\pi}{k}\, g_1(x_1)\, g_2(x_2) \,(r^-_{12}\, \theta (x_{12}) + r^+_{12}\, \theta (x_{21}) )\quad{\rm for}\quad |x_{12}| <2\pi\qquad
\lb{PBg}
\ea
where $r_{12}$ is some skewsymmetric ($r_{12}= - r_{21}$) solution of the modified Yang-Baxter equation (YBE)
\be
[[ r ]]_{123} := [r_{12}, r_{13} ] + [r_{12}, r_{23} ] + [r_{13}, r_{23} ] = [C_{12}, C_{23} ]\ .
\lb{mCYBE}
\ee
It follows from (\ref{mCYBE}) that $r_{12}^\pm = r_{12}  \pm C_{12}$ both solve the ordinary YBE, $[[ r^\pm ]]_{123} = 0\,.$
Such pairs $r^\pm$ (note that $r^+_{12} - r^-_{12} = 2\,C_{12}$) provide a {\em factorization} of
${\cal G} = {\cal G}^+ + {\cal G}^- $ where ${\cal G}^\pm$ are Lie subalgebras of ${\cal G}$,
i.e. any $X\in {\cal G}$ can be represented uniquely as $X = X^+ - X^-\,,\ X^\pm = \frac{1}{2}\,r^\pm X \in {\cal G}^\pm$
so that $r X = X^+ + X^-$ \cite{S-T-S, RS}. This factorization can be lifted locally to the group.

The modified YBE (\ref{mCYBE}) however has no solutions for ${\cal G}$ compact \cite{CGR} so that working with constant classical $r$-matrices
requires complexification. For example, if ${\cal G}$ is the compact form of a complex semisimple Lie algebra and $e_{\pm\a}$ are the raising
and lowering step operators, respectively, corresponding to the positive and negative roots in a Cartan-Weyl basis of the latter,
the so called {\em standard} solution of (\ref{mCYBE}) has the form
\be
r_{12} = \sum_{\a >0} ( (e_\a)_1 (e_{-\a})_2 - (e_{-\a})_1 (e_\a)_2 )\ .
\lb{rstandard}
\ee
The corresponding factorization of the monodromy matrix
\be
M\, =\, M_+\, M_-^{-1}\ ,\quad M_\pm \in B_\pm\ ,\quad {\rm diag}\, M_+  ={\rm diag}\, M_-^{-1} =: D\ ,
\lb{M+-}
\ee
where $M_\pm$ belong to the Borel subgroups $B_\pm$ of the complex group, is a modification of the Gauss decomposition
valid on a local dense neighbourhood of the unit element. For $G=SU(n)\,,$ $\det D = 1$ and $\,B_\pm\subset SL(n)$ are just the groups of complex
unimodular upper and lower triangular matrices. As the Borel algebras are solvable, $\theta (M_\pm)=0\,;$ using this fact, one can prove directly that
\be
\rho (M)\, =\, {\rm tr}\, (M_+^{-1} d M_+ \wedge M_-^{-1} d M_- )
\lb{rho}
\ee
satisfies (\ref{WZM}). Then the $r$-matrix (\ref{rstandard}) is the one appearing in (\ref{PBg}) after inverting the
chiral symplectic form $\Omega (g , M)$ that involves $\rho(M)$ (\ref{rho}) \cite{FG1}.

\smallskip

The following comment is in order. The dynamical $r$-matrix $r_{12}(p)$ in the Bloch waves' PB (\ref{PBBW}) is essentially fixed,
the only freedom being in its diagonal entries while the nontrivial off-diagonal ones,
\be
r (p)^{j\ell}_{~\ell j} = -  i \frac{\pi}{k}\,{\rm cot} \left(\frac{\pi}{k} p_{j\ell}\right)
\quad{\rm for}\quad j\ne\ell
\lb{dyn-r-matr}
\ee
do not depend on further conventions \cite{BDF, BFP}. By contrast, the $r$-matrix entering the chiral field's PB is
to a large extent a matter of choice; even in the particular (monodromy independent) case (\ref{PBg}) it could be
any solution of (\ref{mCYBE}). (This can be achieved by properly choosing the $2$-form $\rho (M)$ subject to (\ref{WZM}) while
in the Bloch waves' case one can only vary the aforementioned {\em closed} form $\rho (M_p)$). We shall deal here with the
$r$-matrix (\ref{rstandard}) which is the quasiclassical limit of the Drinfeld-Jimbo quantum $R$-matrix for $U_q(s\ell(n))\,.$

The PB (\ref{PBBW}) and (\ref{PBg}) are invariant with respect to chiral periodic {\em left} shifts
(half of the invariance inherited from the $2D$ field), a symmetry generated by the chiral Noether current $j(x)\,.$
Both fields $g(x)\,$ and $u(x)\,$ are related to $j(x)$ by the {\em classical} Knizhnik-Zamolodchikov (KZ) equation
\be
i k \frac{d g}{d x} (x) = j (x)\, g (x) \ ,\qquad i k \frac{d u}{d x} (x) = j (x)\, u (x)\ .
\lb{clKZ}
\ee
The transformation properties of $g(x)$ and $u(x)$ with respect to {\em right} shifts however differ. In particular, the right
symmetry of (\ref{PBg}) $g(x) \to g(x)\, S$ requires the PB of the (constant) transformation matrices $S\in G$ to be nontrivial:
\be
\{ S_1 , S_2 \} = \frac{\pi}{k}\, [r_{12} , S_1 S_2 ]\ .
\lb{SklB}
\ee
The {\em Sklyanin bracket} (\ref{SklB}) indicates that this symmetry is of {\em Lie-Poisson} type \cite{D1, S-T-S}.

The solutions of (\ref{clKZ}) are proportional to the path ordered exponential of $j(x)$ and so can only differ by their initial values, hence
\be
g(x) = u(x)\, a\quad\Rightarrow\quad  a\, M = M_p\, a\ .
\lb{gua}
\ee
The introduction of the {\em chiral zero mode} matrix $a=(a^i_\a)\,$ makes it possible to present the symplectic form
$\Omega (g,M)$ as a sum of the ones for the Bloch waves and the zero modes sharing the {\em same\,}
diagonal monodromy $M_p$ \cite{FG1, FHT6}. It is advantageous to first extend the phase space by introducing two
independent $M_p$ and impose their equality as a first order constraint at a later stage.

One finds the following PB for $a^i_\a$ and $p_j\,$ (subject to (\ref{bary})):
\be
\fl
\quad
\{ p_i , p_j \} = 0\ ,\quad\{ a_{\alpha}^i , p_j \} = i\, (\delta_j^i - \frac{1}{n} )\, a_{\alpha}^i\ ,
\quad \{ a_1 , a_2 \} = r_{12} (p) \, a_1 \, a_2 - \frac{\pi}{k} \, a_1 \, a_2 \, r_{12}\ .
\lb{PBa}
\ee
Note that in this setting the complexification related to the choice of a constant $r$-matrix is attributed entirely to the zero modes.
We shall also display the PB of the monodromy matrix (related to $M_p$ by (\ref{gua})),
\be
\fl
\qquad
\{ M_1 , M_2 \} = \frac{\pi}{k}\ (M_1 r_{12}^-\, M_2 + M_2\, r_{12}^+ M_1\, - M_1 M_2\, r_{12} - r_{12} M_1 M_2 )\ ,
\lb{PBMM}
\ee
those of its Gauss components,
\be
\fl
\qquad
\{ M_{\pm 1} , M_{\pm 2} \} = \frac{\pi}{k}\, [ M_{\pm 1} M_{\pm 2} , r_{12}\, ] \ ,\quad
\{ M_{\pm 1} , M_{\mp 2} \} = \frac{\pi}{k}\, [ M_{\pm 1} M_{\mp 2} , r^\pm_{12}\, ]\ \
\lb{PBMpm}
\ee
as well as the corresponding ones with the zero modes:
\be
\fl
\qquad
\{ M_1 , a_2 \} = \frac{\pi}{k}\, a_2 (r^+_{12} M_1 - M_1 r^-_{12} )\ ,\qquad
\{ M_{\pm 1} , a_2\} = \frac{\pi}{k}\, a_2\, r^\pm_{12}\, M_{\pm 1}\ .
\lb{PBMa}
\ee

The canonical quantization of the chiral model prescribes commutators in place of the linear PB while quadratic ones
like (\ref{PBBW}),
(\ref{PBg}), (\ref{PBa}) or (\ref{SklB}) give rise to quantum $R$-matrix {\em exchange relations} with appropriate quasiclassical and symmetry
properties. In particular, the zero modes satisfy
\ba
&&R_{12} (p) \, a_1 \, a_2 = a_2 \, a_1 \, R_{12}\quad\Leftrightarrow\quad
{\hat R}_{12} (p) \, a_1 \, a_2 = a_1 \, a_2 \, {\hat R}_{12}\ ,\nn\\
&&{\hat R}_{12} := P_{12} R_{12}\ ,\quad {\hat R}_{12} (p) := P_{12} R_{12} (p)\ ,
\lb{exa}
\ea
where $P_{12}$ is the permutation matrix. Here the constant (Drinfeld-Jimbo) quantum $R$-matrix is given by
\be
\fl
\qquad
q^{-\frac{1}{n}}\, {\hat R}^{\a\b}_{~\a'\b'} = \d^\b_{\a'}\d^\a_{\b'} + (q^{-1}- q^{\epsilon_{\b\a}} )\, \d^\a_{\a'}\d^\b_{\b'} \ ,
\quad \epsilon_{\a\b} =
\left\{
\begin{array}{ll}
\, -1\,, \ &\a < \b\\
\, 0\,, \ &\a = \b\\
\, 1\,,\ &\a > \b
\end{array}
\right.
\lb{R}
\ee
with
\be
q = e^{-i\frac{\pi}{h}}\ ,\quad h = k+n\,
\lb{q}
\ee
and the quantum {\em dynamical} $R$-matrix \cite{I2, EV2} by
\be
\lb{Rp-ice}
q^{- \frac{1}{n}}\,{\hat R} (p)^{ij}_{~i'j'}\, =\,a_{ij}(p)\, \delta^i_{j'} \delta^j_{i'} + b_{ij}(p)\, \delta^i_{i'} \delta^j_{j'}
\ee
where $a_{ii}(p) = q^{-1}\,,\  b_{ii}(p) = 0\,$ while, for $i\ne j\,,$
\be
\fl
\qquad
a_{ij}(p) = q^{- \a_{ij}(p_{ij})} \,\frac{[p_{ij}-1]}{[p_{ij}]} \quad (\,\a_{ij}(p_{ij}) = - \a_{ji}(p_{ji})\,) \ ,\quad
b_{ij}(p) = \frac{q^{-p_{ij}}}{[p_{ij}]}\ \
\lb{canRp}
\ee
(see \cite{HIOPT, FHIOPT}). We denote $p_i - p_j = p_{ij}\,,$ and the quantum bracket is defined as
$[p] = \frac{q^p - q^{-p}}{q-q^{-1}}\,.$
The {\em operators} $q^{\pm p_j}\,,\ j=1,\dots ,n\ $ ($q^{\pm p_j} q^{\mp p_j} = \id$)
form a commutative set, $q^{p_i} q^{p_j} = q^{p_j} q^{p_i}\,,$ obeying also
\be
\prod_{j=1}^n q^{p_j} = 1\ ,\qquad q^{p_j}\, a_{\alpha}^i = a^i_\a\, q^{p_j + \d^i_j - \frac{1}{n}} \ .
\lb{q^p}
\ee
As the quantum $R$-matrix solves the quantum Yang-Baxter equation (YBE), $\hat R$ obeys the braid relations:
\ba
&&R_{12}\, R_{13}\, R_{23}\, = \, R_{23}\, R_{13}\, R_{12} \qquad\Rightarrow\qquad
{\hat R}_i\, {\hat R}_{i+1}\, {\hat R}_i\, =\, {\hat R}_{i+1}\, {\hat R}_i\, {\hat R}_{i+1}\ ,\nn\\
&&[ {\hat R}_i , {\hat R}_j ] = 0\quad{\rm for}\quad |i-j|\ge 2\quad\, {\rm where}\quad
{\hat R}_i \equiv {\hat R}_{i\,i+1}\ .
\lb{YBE}
\ea
The matrix $R_{12}(p)$ obeying the quantum {\em dynamical} YBE gives rise to another representation of the braid group
\cite{HIOPT}.

The PB of the monodromy matrix and its Gauss components (\ref{PBMM}), (\ref{PBMpm}) and (\ref{PBMa}) are replaced by the exchange relations
\be
\fl
\qquad
R_{12}\, M_2\, R_{21}\, M_1 = M_1\, R_{12}\, M_2\, R_{21}\quad\Leftrightarrow\quad
{\hat R}_{12}\, M_2\, {\hat R}_{12}\, M_2 = M_2\, {\hat R}_{12}\, M_2\, {\hat R}_{12}\ ,\ \
\lb{exM}
\ee
\ba
\fl\qquad
R_{12} M_{\pm 2} M_{\pm 1} = M_{\pm 1} M_{\pm 2} R_{12}\ ,\quad\qquad\,
R_{12} M_{+2} M_{-1} = M_{-1} M_{+2} R_{12}\qquad\Leftrightarrow\nn\\
\fl\qquad{\hat R}_{12}\, M_{\pm 2}\, M_{\pm 1}\, = \, M_{\pm 2}\, M_{\pm 1}\, {\hat R}_{12}\ ,\qquad
M_{- 2}^{-1}\, {\hat R}_{12}\, M_{+2}\, = \, M_{+1}\, {\hat R}_{12}\, M_{-1}^{-1}\qquad\quad
\lb{exMpm}
\ea
and
\be
\fl
\qquad
M_1\, a_2 = a_2 \, R_{21} M_1 R_{12} \equiv a_2\, {\hat R}_{12} M_2 {\hat R}_{12}\ ,\quad
M_{\pm 2}\, a_1 = a_1\, R_{12}^\mp M_{\pm 2}\ ,
\lb{aMpm}
\ee
respectively, for $R_{12}^- = R_{12}\,,\ R_{12}^+ = R_{21}^{-1}\,.$

The quasiclassical correspondence, requiring the leading term in the small $\hbar$ expansion of the commutator $[ A , B ]$
of two quantum dynamical variables to reproduce the PB $i \hbar\, \{ A , B \}$ of their classical counterparts,
is confirmed by the $\frac{1}{k} \to 0$ asymptotics of the exchange relations listed above. To this end, one
uses the expansions
\be
R_{12} = \id_{12} - i\frac{\pi}{k}\, r^-_{12} +{\cal O}(\frac{1}{k^2})\ ,\qquad
R_{21} = \id_{12} + i\frac{\pi}{k}\, r^+_{12} +{\cal O}(\frac{1}{k^2})
\lb{Rr}
\ee
and assumes that the terms of the type $\frac{p_{ij}}{k}$ arising from the dynamical $R$-matrix (\ref{Rp-ice}), (\ref{canRp})
in (\ref{exa}) have a finite quasiclassical limit \cite{FHT6}.

\smallskip

The analytic picture exchange relations for the chiral field
\be
P_{12}\, g_1(z_1)\, g_2(z_2) =\, \stackrel{\curvearrowright}{g_1(z_2)\, g_2(z_1)} {\hat R}_{12}\ ,\quad z = e^{ix}
\lb{Exg}
\ee
for $z_{12} \stackrel{\curvearrowright}{\rightarrow} z_{21} = e^{-i\pi} z_{12}$ \cite{FHIOPT} involve the matrix $\hat R\,,$ while
its conformal properties and twisted periodicity (cf. (\ref{cM})) imply the univalence relation
\be
e^{2\pi iL_0} \, g (z) \, e^{-2\pi i L_0} \equiv e^{2\pi i \Delta} \, g(e^{2\pi i}\,z) = g(z) \, M
\lb{LgM}
\ee
where $L_0$ is the Virasoro operator generating dilations of $z$ and
\be
\Delta = \frac{C_2(\pi_f)}{2h}=\frac{n^2-1}{2nh}
\lb{Delta}
\ee
the conformal dimension of $g(z)$ (here $C_2(\pi_f)$ is the value of the quadratic Casimir operator in the defining
$n$-dimensional representation of $s\ell(n)$).
The current-field commutation relations assume the form
\be
\fl
\quad
[ j_m^a , g (z) ] = - z^m\, T^a \, g (z)\quad{\rm for}\quad j(z) = j^a(z)\, T_a\ ,\quad
j^a(z) = \sum_m j^a_m\, z^{-m-1}
\lb{jg}
\ee
i.e., $g(z)$ is a primary field with respect to the current algebra.

\section{The chiral state space}
\label{sec:3}

\setcounter{equation}{0}
\renewcommand\theequation{\thesection.\arabic{equation}}

We will assume that the state space ${\cal H}$ of the quantized chiral WZNW model is a vacuum (lowest energy)
representation of the exchange algebra (\ref{Exg}) where the quantized chiral field $g(z)$ splits as in (\ref{gua}):
\be
g^A_\a(z) = u^A_i(z) \otimes a^i_\a\ .
\lb{guaq}
\ee
Here the field $u(z) = (u^A_i(z))$ has diagonal monodromy, and introducing the three types of indices
(capital, latin and greek letters) reflects the different nature of the corresponding transformation
properties (of group, "dynamical" and quantum group type, respectively) of the involved objects.
As the zero modes commute with the current, the conformal properties of $u(z)$ and those of the chiral field
coincide. The following chain of relations illustrates how this works in the case of (\ref{LgM}):
\ba
&&e^{2\pi i L_0} u (z)\, e^{- 2\pi i L_0}\otimes\, a = e^{2\pi i \Delta} u(e^{2\pi i}\,z) \otimes\, a =\nn\\
&&= M_p\, u(z) \otimes\, a = u(z) \otimes\, M_p\, a = u(z) \otimes\, a\, M\ .
\lb{LuaM}
\ea
Note that the zero mode matrix $a$ "inherits" the diagonal monodromy of $u(z)$ (the fourth equality above);
this requirement is the quantum counterpart of the fact that, classically, the symplectic forms of the
zero modes and the Bloch waves are not completely independent but share the same $M_p\,.$
Note that, due to the identical exchange relations of $u_i$ and $a^i$ with $p_\ell\,,$
\be
p_\ell \, u^A_i (z)  \, =\, u^A_i (z)\, (p_\ell + \d^i_\ell - \frac{1}{n})\ ,\qquad
p_\ell \, a^i_\a\, =\, a^i_\a\, (p_\ell + \d^i_\ell - \frac{1}{n})\ ,
\lb{gCVO}
\ee
it is important that in the quantum case $M_p$ appears, as a matrix of operators, from the {\em left} side of $u(z)$
(see the third equality in (\ref{LuaM})).
Assuming that ${\cal H}$ is generated from the vacuum vector $\vac$ by polynomials in $g(z)$ (\ref{guaq})
(and its derivatives) implies the following structure of the chiral state space:
\be
\lb{Hspace}
{\cal H}\, =\, \bigoplus_p\,{\cal H}_p\otimes {\cal F}_p \ .
\ee
Here both ${\cal H}_p$ and ${\cal F}_p\,$ are eigenspaces corresponding to the same eigenvalues
of the collection of commuting operators $p = (p_1 ,\dots , p_n)\,$ (to not overburden notation, we will use
in this case the same letter for operators and their eigenvalues; the meaning should be clear from the context).
The (discrete) joint spectrum of $p$
is generated from the vacuum value $p^{(0)}$ according to the rules implied by (\ref{gCVO}).
We will assume that the vacuum vector is unique so that ${\cal H}_{p^{(0)}} = {\mathbb C} \vac$ is one dimensional.

As the current $j(z)$ commutes with $p\,,$ the spaces ${\cal H}_p\,$ are invariant with respect to the
(conformal) current algebra, the corresponding representations being
not necessarily irreducible. The (columns of) $(u^A_i(z))$ act as elementary intertwining operators
analogous to the chiral vertex operators in the axiomatic approach to the model.

Similarly, each ${\cal F}_p$ is a quantum group representation space.
To see this, one notes that the monodromy $M$ as well as its Gauss components also commute with $p$
and further, that the exchange relations (\ref{exMpm}) supplemented by
\be
\prod_{\a=1}^n d_\a = 1\qquad{\rm for}\qquad d_\a := (M_+)^\a_{~\a} = (M_-^{-1})^\a_{~\a}\ ,\quad \a =1,\dots ,n \ ,
\lb{prod-d1}
\ee
together with the natural coalgebraic structure assuming $\Delta (1) = 1\otimes 1$ and
\be
\fl
\quad
\Delta ((M_\pm)^\a_{~\b}) = ( M_\pm )^\a_{~\s}\otimes (M_\pm)^\s_{~\b}\ ,\quad
\varepsilon ((M_\pm)^\a_{~\b}) = \d^\a_\b\ ,\quad S ((M_\pm)^\a_{~\b}) =  (M_\pm^{-1})^\a_{~\b}
\lb{Hopf-FRT}
\ee
($\Delta\,,\ \varepsilon$ and $S$ being the coproduct, counit and the antipode, respectively)
define a Hopf algebra equivalent to an $n$-fold cover $U_q$ of $U_q(s\ell(n))$ \cite{FRT, HF2}.
In particular, it follows from (\ref{Hopf-FRT}) and the triangularity of the matrices $M_\pm\,$ that their
diagonal elements are necessarily group-like, i.e.
$\Delta (d_\a^{\pm 1}) = d_\a^{\pm 1}\otimes\, d_\a^{\pm 1}\,.$
On the other hand, relations (\ref{exMpm}) (with ${\hat R}_{12}$ given by (\ref{R})) show that $\{ d_\a \}$ commute
and can be expressed in terms of Cartan generators $\{ k_i \}\,$ corresponding to the fundamental weights\footnote{The Cartan
generators $K_i$ of $U_q(s\ell(n))$ corresponding to the simple (co-)roots are given by $K_i = k_{i-1}^{-1} k_i^2 k_{i+1}^{-1}$
and an inverse formula expressing $k_i$ in terms of $K_i$ would involve "$n$-th roots" of the latter. This explains the term
"$n$-fold cover" \cite{HF2} characterizing the Hopf algebra $U_q$ (called the "simply-connected rational form" in \cite{CP}).}:
\be
\fl
\quad
d_\a = k_{\a-1} k^{-1}_\a\quad (\,k_0 = k_n = 1\,)\qquad\Leftrightarrow\qquad k_i = \prod_{\ell =1}^i d_\ell^{-1}\ ,\quad i=1,\dots ,n-1\ .
\lb{dkk}
\ee
Further, the $n-1$ non-zero next-to-diagonal entries of $M_\pm$ are related to the step operators (lowering and
raising, respectively) and the other non-zero entries, to the remaining Cartan-Weyl basis elements.

\smallskip

\noindent
{\bf Remark 3.1~}
The general structure (\ref{Hspace}) of ${\cal H}$ reminds the one predicted by local quantum field theory \cite{Haag}.
The spaces ${\cal H}_p$ correspond to the superselection sectors of the algebra of observables
(generated in our case by the current) and ${\cal F}_p\,,$ to the finite-dimensional representations of the gauge
(internal) symmetry which leaves the observables invariant. While in space-time dimension $D\ge 4$ the gauge
symmetry is necessarily a compact group (Doplicher-Roberts' theorem \cite{DR}), here this role is played by the
quantum group $U_q\,$ and the permutational Bose-Fermi alternative is replaced by a (nonabelian)
braid group statistics, cf. (\ref{Exg}).

The mere fact that the labels $p$ are common for both ${\cal H}_p$ and ${\cal F}_p$ assumes that they provide
(at least a partial) characterization of both representation spaces. This is not completely trivial since the
represented algebras are of different nature. In the case at hand the "superselection charges"
$p = (p_1 ,\dots , p_n)$ are related both to the $n-1$ independent Casimir operators of $su(n)$ that
label the representations of the affine algebra ${\widehat{su}}(n)_k$ and to the deformed Casimirs of
(a quotient \cite{FHT7} of) the Hopf algebra $U_q\,.$

\smallskip

As the deformation parameter is a root of unity, the dynamical $R$-matrix (\ref{Rp-ice}) is singular for $p_{ij} = n h\,,\ n\in {\mathbb Z}\,,$
and so the exchange relations (\ref{exa}) are ill defined on ${\cal F}\,.$
However, getting rid of the dangerous denominators and using the identity $[p-1]-q^{\pm 1} [p] = -\, q^{\pm p}\,,$
we obtain (with $\a_{ij}(p_{ij})$ in (\ref{canRp}) set to zero) the following set of relations that always make sense:
\ba
&&a^j_\alpha a^i_\beta\, [p_{ij}-1] = a^i_\beta\, a^j_\alpha\, [p_{ij}] -\,
a^i_\alpha a^j_\beta \, q^{{\epsilon}_{\beta\alpha}p_{ij}} \quad (\,{\rm for}\quad i\ne j \quad {\rm and}\quad\alpha\ne\beta )\ ,\nn\\
&&[a^j_\alpha , a^i_\alpha ] = 0\ ,\qquad a^i_\alpha a^i_\beta = q^{{\epsilon}_{\alpha\beta}}\, a^i_\beta\, a^i_\alpha\ ,\qquad i,j=1,\dots, n\ .
\lb{aa2}
\ea

\section{The zero modes' quantum determinant}
\label{sec:4}

Following \cite{HIOPT, FHIOPT}, we will supply the algebra generated by $\{ a^i_\a \}$ and $\{ q^{\pm p_j} \}$
satisfying (\ref{q^p}) and the (quadratic in the zero modes) exchange relations (\ref{aa2})
with an additional $n$-linear relation for the {\em quantum determinant} $\,{\det}_q (a)\,.$
The fact that both the constant and the dynamical $R$-matrix are of Hecke type{\footnote{This is a
special property of the quantum deformation of $s\ell(n)\simeq A_{n-1}$ \cite{FRT}.},
\be
(q^{-\frac{1}{n}}\,\hat R - q^{-1} ) (q^{-\frac{1}{n}}\,\hat R + q) = 0 = (q^{-\frac{1}{n}}\,{\hat R}(p) - q^{-1} ) (q^{-\frac{1}{n}}\,{\hat R}(p) + q)
\lb{Hecke}
\ee
allows to introduce elementary constant and dynamical {\em quantum antisymmetrizers} by
\be
q^{-\frac{1}{n}}\,{\hat R}_{12} = q^{-1}\id - A_{12}
\lb{q-anti}
\ee
(and similarly for the dynamical one). Higher antisymmetrizers $A_{1j}$ can be defined inductively from (\ref{q-anti})
and $A_{11} = \id\,$ \cite{HIOPT}.
One notices that, for $q$ given by (\ref{q}), $A_{1 n+1} = 0$ and $A_{1 n}$ is proportional to a rank $1$ projector
(same as in the undeformed case). As a result, $A_{1n}$ is of the form
\be
(A_{1n})^{\alpha_1 \ldots \alpha_n}_{~\beta_1 \ldots \beta_n} =
\varepsilon^{\alpha_1 \ldots \alpha_n}\, \varepsilon_{\beta_1 \ldots \beta_n}
\lb{A1n}
\ee
where the $\varepsilon$-tensors, the deformed analogs of the "ordinary" fully antisymmetric tensors
of rang $n$, satisfy the equations
\ba
&&{\hat R}^{\alpha_i \alpha_{i+1}}_{~\s_i \s_{i+1}}\, \varepsilon^{\alpha_1\dots \s_i \s_{i+1}\dots \alpha_n}=
- q^{1+\frac{1}{n}}\,\e^{\a_1 \dots \a_n}\ ,\nn\\
&&\varepsilon_{\alpha_1\dots \s_i \s_{i+1}\dots \alpha_n}\, {\hat R}_{~\alpha_i \alpha_{i+1}}^{\s_i \s_{i+1}}=
- q^{1+\frac{1}{n}}\,\e_{\a_1 \dots \a_n}\ ,\quad i=1,\dots, n-1\ .\qquad
\lb{eqs-epsR}
\ea
As one can verify directly, by using the explicit form of ${\hat R}_{12}$ (\ref{R}), Eqs.(\ref{eqs-epsR}) imply
in particular that the constant $\varepsilon$-tensors vanish if some of the indices coincide.
After fixing conveniently the intrinsic normalization freedom, their non-zero components are explicitly given by
\begin{equation}
\label{q-eps}
\varepsilon^{\alpha_1 \ldots \alpha_n} = \varepsilon_{\alpha_1 \ldots \alpha_n} = q^{- \frac{n(n-1)}{4}} \, (-q)^{\ell (\sigma)}
\qquad \mbox{for} \quad \sigma = \left( {n\ \ldots\ 1}\atop{~\alpha_1 \ldots ~\alpha_n} \right)
\in {\mathcal S}_n\ ,
\end{equation}
where ${\mathcal S}_n$ is the symmetric group of $n$ objects and $\ell (\sigma)$ is the length of the permutation $\s\,.$

The dynamical $\epsilon$-tensors can be found by a similar procedure.
We will choose the one with lower indices to be equal to its undeformed counterpart,
\be
\epsilon_{i_1 \ldots i_n} \, (p) = \epsilon_{i_1 \ldots i_n} \qquad ( \epsilon_{n\, n-1 \ldots 1} = 1 )
\lb{ep-low}
\ee
in which case the non-zero components of that with upper indices are
\be
\epsilon^{i_1 \ldots i_n} \, (p) = \frac{(-1)^{\frac{n(n-1)}{2}}}{{\cal D}_q(p)}\,
\prod_{1 \leq \mu < \nu \leq n} [p_{i_{\mu} i_{\nu}} - 1]\ ,\qquad
{\cal D}_q(p) := \prod_{i<j} [p_{ij}]\ .
\lb{eps-Dqp}
\ee
One can verify that both tensors obey the normalization condition
\be
\varepsilon^{\alpha_1 \ldots \alpha_n} \varepsilon_{\alpha_1 \ldots \alpha_n} = [n]! =
\epsilon^{i_1 \ldots i_n} \, (p)\, \epsilon_{i_1 \ldots i_n} \, (p)\ .
\lb{norm-e}
\ee
Now the quantum determinant of the matrix $a$ is defined as
\be
\lb{Dqa}
{\det}_q (a) := \frac{1}{[n]!} \, \epsilon_{i_1 \ldots i_n} (p) \, a_{\alpha_1}^{i_1} \ldots a_{\alpha_n}^{i_n} \,
\varepsilon^{\alpha_1 \ldots \alpha_n}\ ,\quad [n]! = [n] [n-1]\dots 1\ .
\ee
The following two facts \cite{HIOPT} will be of major importance for what follows:

\noindent
1) the product $a_1 \ldots a_n$ intertwines between the constant and dynamical epsilon-tensors,
\ba
&&\epsilon_{i_1 \ldots i_n} (p) \, a_{\alpha_1}^{i_1} \ldots a_{\alpha_n}^{i_n} =
{\det}_q (a) \, \varepsilon_{\alpha_1 \ldots \alpha_n}\ ,\nn\\
&&a_{\alpha_1}^{i_1} \ldots a_{\alpha_n}^{i_n} \,
\varepsilon^{\alpha_1 \ldots \alpha_n} = \epsilon^{i_1 \ldots i_n} \, (p) \, {\det}_q (a)\ ;
\lb{det-intertw}
\ea
2) the ratio $\frac{{\det}_q(a)}{{\cal D}_q(p)}$ is central for the algebra generated by $\{ a^i_\a \}$ and $\{ q^{\pm p_j} \}\,,$
\be
\fl
\qquad
[ q^{p_i} , {\det}_q (a) ] = 0 \ (\, = [ q^{p_i} , {\cal D}_q(p) ]\,)\ ,\quad [\, \frac{{\det}_q(a)}{{\cal D}_q(p)}\, ,\, a^i_\a ]= 0\ ,\quad i,\a=1,\dots ,n
\lb{qcent}
\ee
so it is reasonable to postulate that the quantum determinant of zero modes' matrix $a$ is {\em equal}
(not  to $1$ but) to ${\cal D}_q(p)$:
\be
{\det}_q(a) = {\cal D}_q(p) \equiv \prod_{1\le i<j\le n} [p_{ij}]\ .
\lb{Dqap}
\ee

\section{Quantum prefactors of the monodromy matrices}
\label{sec:5}

\setcounter{equation}{0}
\renewcommand\theequation{\thesection.\arabic{equation}}

Taking the limit $z \to 0$ in (\ref{LuaM}) (which is possible due to energy positivity and is at
the heart of the "operator-state correspondence"), one ends up with a set of conditions which
only involve operators acting solely on the zero modes' space ${\cal F} := \oplus_p {\cal F}_p\,,$
\be
e^{2\pi i \Delta} a^i_\a \vac\  \equiv q^{\frac{1}{n}-n} a^i_\a \vac \ = (M_p)^j_i\, a^j_\a \vac
= a^i_\b\, M^\b_{~\a} \vac
\lb{main}
\ee
(we have taken into account (\ref{Delta})). One thus obtains, in particular,
\be
M^\a_{~\b} \vac
= q^{- C_2(\pi_f)} \d^\a_\b \vac
= q^{\frac{1}{n} - n}\, \d^\a_\b \vac
\lb{M0}
\ee
i.e., the vacuum is annihilated by the off-diagonal elements of $M\,$ and is a common eigenvector of the diagonal ones, corresponding to the (common) eigenvalue $q^{\frac{1}{n} - n}\,.$
On the other hand, the parametrization of $M_\pm$ in terms of $U_q$ generators discussed above makes it obvious that the quantum group invariance of the vacuum is equivalent to a similar
condition for $M_\pm\,,$
\be
\fl
\qquad
X \vac = \varepsilon (X) \vac\quad \forall X\in U_q\quad\Leftrightarrow\quad
(M_\pm)^\a_{~\b} \vac = \varepsilon ((M_\pm)^\a_{~\b} ) \vac = \d^\a_\b \vac
\lb{Uqvac}
\ee
where $\varepsilon (X)$ is the counit (\ref{Hopf-FRT}).
Comparing (\ref{M0}) and (\ref{Uqvac}), we conclude that the factorization of the quantum monodromy matrix
$M$ in upper and lower triangular Gauss components of the type (\ref{M+-}) should be modified to
\be
\fl\qquad
M=q^{\frac{1}{n} - n} M_+ M_-^{-1} \qquad (\, {\rm diag}\, M_+ = {\rm diag}\, M_-^{-1} = D\ ,\quad \det D = 1\,)\ .
\lb{M+-q}
\ee

It is natural to expect that the quantum {\em diagonal} matrix $M_p$ has to be modified accordingly.
The striking point is that, although the intertwining property of the zero modes' matrix in (\ref{LuaM}) is the
same as in (\ref{gua}), $M_p$ gets a quantum prefactor {\em different} from that of $M\,.$ More precisely,
the classical parametrization (\ref{uMp}), (\ref{bary}) amounts to $(M_p)_j^i = {q_{cl}}^{ - 2 p_i} \, \delta_j^i$
with $q_{cl} = e^{-i\frac{\pi}{k}}$ while in the quantum case $q$ is given by (\ref{q}); the analysis shows
that, apart from this (well known) replacement of the level $k$ by the height $h$, the correct expression
for the quantum  diagonal matrix should be
\be
\lb{Mpq}
(M_p)_j^i = q^{ - 2 p_i + 1-\frac{1}{n}} \, \delta_j^i\ .
\ee
The field-theoretic arguments in favor of this choice will be spelled out below.
Plugging (\ref{Mpq}) and (\ref{M0}) into (\ref{main}) and using (\ref{gCVO}), we obtain
\be
q^{\frac{1}{n}-n}\, a^i_\a\, {\mid 0 \rangle} = a^i_\a\, q^{-2p_i - 1 + \frac{1}{n}}\,
{\mid 0 \rangle}\quad{\rm i.e.,}\quad
a^i_\a\, q^{-2p_i}\, {\mid 0 \rangle} = q^{1-n}\, a^i_\a\, {\mid 0 \rangle}\ .
\lb{aqp-vac}
\ee
Eq.(\ref{aqp-vac}) admits the following interpretation.

\noindent
1) The vacuum eigenvalues $p^{(0)}_i$ on ${\mid 0 \rangle}$ are equal to the barycentric coordinates $p_i(\rho)$
of the Weyl vector (the latter, being defined as the half-sum of the positive roots, is also equal
to the sum of the $n-1$ fundamental weights $\L^j$ or, in other words, all its Dynkin labels $\l_j$ are equal to $1$):
\ba
&&\rho := \frac{1}{2} \sum_{\a >0} \a = \sum_{j=1}^{n-1} \L^j\ ,\quad \l_j(\rho) = 1\ ,\quad j=1,\dots, n-1\ ;\nn\\
&&p_i\, {\mid 0 \rangle} = p^{(0)}_i\,{\mid 0 \rangle}\ ,\quad
p^{(0)}_i = p_i(\rho) = \frac{n+1}{2} - i\ ,\quad i=1,\dots, n\qquad
\lb{vac-Weyl}
\ea
so that, in particular, $q^{-2p^{(0)}_1} = q^{1-n}$;

\noindent
2) All operators $a^i_\a$ with $i\ne 1$ annihilate the vacuum vector:
\be
a^i_\a \,{\mid 0 \rangle} = 0\quad{\rm for}\quad i\ge 2\ .
\lb{a2.n}
\ee

\noindent
These two assumptions guarantee the validity of (\ref{aqp-vac}).

\smallskip

Eq.(\ref{gCVO}) provides a simple visualization of the action of the operators $a^i_\a$:
for a given $i\,,$ it corresponds to adding a box to the $i$-th line of an $s\ell (n)$-type Young diagram,
the additional condition (\ref{Dqap}) accounting for the triviality of the determinant representation.
Hence, if a homogeneous polynomial ${\cal P}^\L (a)$ is associated to the representation with
highest weight $\L = \sum_{j=1}^{n-1} \l_j \,\L^j\,,$ then the eigenvalues of the operators $p\,$ on
the state \mbox{${\cal P}^\L (a)\! \vac \in {\cal F}$} are the barycentric coordinates of the {\em shifted} weight
$\L + \rho\,$ which can be found from
\be
p_{j j+1} = \l_j +1\ ,\quad j=1,\dots, n-1\ ,\quad \sum_{i = 1}^n p_i = 0\ .
\lb{bary-Dynkin}
\ee
It follows from (\ref{Dqap}) that the determinant of $a$
does not vanish (and is positive) on states for which $\L$ satisfies the integrability conditions for $\widehat{su}(n)_k$
\be
\l_j \in {\Z}_+\ ,\quad \sum_{j=1}^n \l_j \le k\quad \Leftrightarrow\quad p_{j j+1} \in {\mathbb N}\ ,
\quad p_{1n} \le h-1\ .
\lb{integr}
\ee

The operators $u^A_i(z)$ have the same exchange properties with $p_\ell\, $ as $\, a^i_\a\,,$ and a {\em regularized}
determinant also exists in this case\footnote{Work in progress with Ivan Todorov.}. The latter is however proportional
to the {\em inverse power} of ${\cal D}_q(p)$  and so may diverge on states not satisfying the integrability conditions
(\ref{integr}). Thus the field $u(z)$ alone cannot be defined on the space $\oplus_p {\cal H}_p\,$
where the joint spectrum of $p$ is assumed to be infinite. On the other hand, due to the regularizing role of the
zero modes, the chiral field $g(z)$ acting on ${\cal H}$ (\ref{Hspace}) provides a sound logarithmic extension of the
chiral WZNW model \cite{STH, HST, HP, FHT7}. Whether there is a way of truncating, within the context
of canonical quantization described so far, the state space ${\cal H}$ (\ref{Hspace}) to a finite direct sum
containing only the integrable values (\ref{integr}) of $p$ remains an open problem.
If this idea turns out to be correct, singling out the truncated space would be similar in spirit to finding the
physical space of states in a covariantly quantized gauge theory.

\smallskip

After discussing the field-theoretical arguments for the quantum corrections to the monodromy matrices,
we will now turn to the algebraic aspects. From (\ref{LuaM}), one would expect the relation
\be
\lb{detaM}
{\det}_q (M_p\, a) = {\det}_q (a) = {\det}_q (a M)
\ee
to hold for appropriately defined ${\det}_q (M_p\, a)$ and ${\det}_q (a M)\,.$
We will show in the next two sections that (\ref{detaM}) indeed takes place for
the corresponding quantum determinants defined in a natural way. Moreover, the
quantum correction factors allow to retain in the quantum case the classical property of
factorization of the matrix product: ${\det}_q (AB) = {\det}_q (A)\, {\det}_q (B)$.

\section{Quantum determinants involving $M_p$}
\label{sec:6}

\setcounter{equation}{0}
\renewcommand\theequation{\thesection.\arabic{equation}}

We will start with the first relation (\ref{detaM}) ${\det}_q (M_p\, a) = {\det}_q (a)$ by showing
that the non-commutativity of $q^{{p}_j}\,$ and $a^i_\a\,,$
cf. (\ref{q^p}), exactly compensates the additional factors $q^{1-{1\over n}}\,$ coming from $M_p$ (\ref{Mpq})
when computing
\be
{\det}_q (M_p\, a) :=
\frac{1}{[n]!}\, \epsilon_{i_1 \ldots i_n}\, (M_p\, a)_{\a_1}^{i_1} \ldots (M_p\, a)_{\a_n}^{i_n}\, \e^{\a_1 \dots \a_n}\ .
\lb{detqMpa}
\ee
(cf. (\ref{Dqa})). As $M_p$ is diagonal, the computation is very simple. Assume that $i_\mu \ne i_\nu$ for $\mu\ne\nu$
(the non-zero terms in (\ref{detqMpa}) have this property due to the presence of the $\epsilon$-tensor)
so that, in particular, $\prod_{\mu=1}^n\, q^{-2p_{i_\mu}} = \prod_{i=1}^n\, q^{-2p_i} = 1\,.$
We then have
\ba
\fl\quad
(M_p\, a)_{\a_1}^{i_1} \ldots (M_p\, a)_{\a_n}^{i_n}\,= q^{-2{p}_{i_1} +1-{1\over n}}\, a^{i_1}_{\a_1}\,
q^{-2{p}_{i_2} +1-{1\over n}}\, a^{i_2}_{\a_2}\,\ldots
q^{-2{p}_{i_n} +1-{1\over n} }\, a^{i_n}_{\a_n} \, = \nn\\
\fl\quad
= \left(\prod_{i=1}^n q^{-2p_i}\right) \,a^{i_1}_{\a_1} a^{i_2}_{\a_2}\ldots a^{i_n}_{\a_n}\, =\,
a^{i_1}_{\a_1} a^{i_2}_{\a_2}\ldots a^{i_n}_{\a_n} \, \left(\prod_{i=1}^n q^{-2p_i} \right)
= a^{i_1}_{\a_1} a^{i_2}_{\a_2}\ldots a^{i_n}_{\a_n}
\lb{qsum}
\ea
since, moving all $q^{-2{p}_{i_\mu} +1-{1\over n}}$ terms either to the leftmost or to the rightmost position,
we get trivial overall numerical factors \cite{FHIOPT}:
\be
\lb{qsum1}
q^{n(1-{1\over n}) - {2\over n} (1+2+\dots + n-1)}\, =\, 1\, = q^{n(1-{1\over n}) - 2n + {2\over n} (1+2+\dots + n)}\ .
\ee
Hence, defining simply
\be
{\det}_q (M_p) := \prod_{i=1}^n q^{-2p_i} \ \  (\, = 1\, ) \ ,
\lb{detMp}
\ee
we obtain
\be
{\det}_q (M_p\, a) = {\det}_q (M_p)\, {\det}_q (a) = {\det}_q (a)\, {\det}_q (M_p)\ .
\lb{DaDMp}
\ee

\section{The quantum determinants ${\det}_q (M)$ and ${\det}_q (M_\pm)$}
\label{sec:7}

\setcounter{equation}{0}
\renewcommand\theequation{\thesection.\arabic{equation}}

The clue to the second relation (\ref{detaM}) ${\det}_q (a) = {\det}_q (a M)\,,$ is given by the equality
\be
a_1 M_1\, a_2 M_2\, \dots a_n M_n = a_1 \,a_2 \dots a_n \, ({\hat R}_{12} {\hat R}_{23} \dots {\hat R}_{n-1\, n} M_n )^n
\lb{aMn}
\ee

\smallskip

\noindent
(the proof of (\ref{aMn}) as well as that of (\ref{MRn}) will be displayed in the next section). Defining
\be
{\det}_q (a\, M) := \frac{1}{[n]!}\, \epsilon_{i_1 \ldots i_n}\, (a\, M)_{\b_1}^{i_1} \ldots
(a\, M)_{\b_n}^{i_n}\, \e^{\b_1 \dots \b_n}\ ,
\lb{detaM1}
\ee
using (\ref{aMn}) and the first relation (\ref{det-intertw}), we obtain
\be
{\det}_q (a M) = {\det}_q (a)\, {\det}_q (M)
\lb{det-mult}
\ee
with the following expression for the determinant of the monodromy matrix satisfying the reflection equation
(\ref{exM}):

\be
{\det}_q (M) := \frac{1}{[n]!}\,\e_{\a_1 \dots \a_n}\,
\left[ ({\hat R}_{12} {\hat R}_{23} \dots {\hat R}_{n-1\, n} M_n )^n\right]^{\a_1 \dots \a_n}_{~\b_1 \dots \b_n}\,\e^{\b_1\dots \b_n}\ .
\lb{detM}
\ee

\smallskip

\noindent
One can further rearrange (\ref{detM}) in terms of the Gauss components of the monodromy matrix, using
\be
\fl\qquad
({\hat R}_{12} {\hat R}_{23} \dots {\hat R}_{n-1\, n} M_n )^n
= q^{1-n^2} ({\hat R}_{12} \dots {\hat R}_{n-1\, n})^n  M_{+ n}\dots M_{+ 1} M_{- 1}^{-1} \dots M_{- n}^{-1}\ .
\lb{MRn}
\ee

\smallskip

\noindent
The exchange relation ${\hat R}_{12}M_{\pm 2}M_{\pm 1}=M_{\pm 2}M_{\pm 1}{\hat R}_{12}$ (\ref{exMpm}) implies
\be
\lb{AMMA}
A_{1n}\, M_{\pm n}\dots M_{\pm 1} = M_{\pm n}\dots M_{\pm 1}\, A_{1n}
\ee
where $A_{1n}$ is the constant quantum antisymmetrizer (\ref{A1n}).
Eq.(\ref{AMMA}) is in turn equivalent to
\ba
&&\e_{\a_1 \dots \a_n}\,(M_\pm )^{\a_n}_{~\b_n}\dots (M_\pm )^{\a_1}_{~\b_1}  = {\det}_q (M_\pm) \,\e_{\b_1\dots \b_n}\ ,\nn\\
&&(M_\pm )^{\a_n}_{~\b_n}\dots (M_\pm )^{\a_1}_{~\b_1}\,\e^{\b_1\dots \b_n} = {\det}_q (M_\pm) \,\e^{\a_1 \dots \a_n}
\lb{detMpmvar}
\ea
where we define
\be
{\det}_q (M_\pm) :=
\frac{1}{[n]!}\,\e_{\a_1 \dots \a_n}\,(M_\pm )^{\a_n}_{~\b_n}\dots (M_\pm )^{\a_1}_{~\b_1} \,\e^{\b_1\dots \b_n}
\lb{detMpmvar1}
\ee
(to show the equivalence of (\ref{AMMA}) and (\ref{detMpmvar}), just use (\ref{norm-e})).
Due to the triangularity of $M_\pm\,,$ the only nontrivial terms in the sum (\ref{detMpmvar1}) are
the $n!$ products of their (commuting) diagonal elements $d_\a^{\pm 1}$, hence
\be
{\det}_q (M_\pm) = \prod_{\a =1}^n (M_\pm )^{\a}_{~\a} = \prod_{\a =1}^n d_\a^{\pm 1} = 1
\lb{detMpmvar2}
\ee
(cf. (\ref{prod-d1})).
Using the antipode $S$ (\ref{Hopf-FRT}), one derives
\be
{\det}_q (M_\pm^{-1}) = {\det}_q (S(M_\pm)) = \prod_{\a =1}^n d_\a^{\mp 1} = 1\ .
\lb{detM-1}
\ee
Due to (\ref{eqs-epsR}), the $q^{1-n^2}$ prefactor in (\ref{MRn}) is exactly compensated by
\be
\fl\qquad
\e_{\a_1 \dots \a_n}\!\left[({\hat R}_{12} {\hat R}_{23} \dots {\hat R}_{n-1\, n} )^n\right]^{\a_1 \dots \a_n}_{~\b_1 \dots \b_n}
= (- q^{1+\frac{1}{n}})^{(n-1)n}\,\e_{\b_1 \dots \b_n}
= q^{n^2-1}\,\e_{\b_1 \dots \b_n}\ .
\lb{epsRij}
\ee
From (\ref{detM}), (\ref{MRn}), (\ref{epsRij}) and (\ref{detMpmvar}), (\ref{detM-1}) we finally obtain
\be
{\det}_q (M) = {\det}_q (M_+) \,.\, {\det}_q (M_-^{-1})\, = 1\ .
\lb{MMMpm}
\ee
Eqs. (\ref{det-mult}) and (\ref{MMMpm}) validate the second relation (\ref{detaM}).

\section{Proofs of two important identities}
\label{sec:8}

Here we shall provide proofs of the two relations (\ref{aMn}) and (\ref{MRn}) which play a crucial role
in the derivation of the relations involving the determinants of the monodromy matrix $M$ (satisfying
the reflection equation (\ref{exM})) and its Gauss components $M_\pm$ (subject to the
exchange relations (\ref{exMpm})).

\smallskip

The {\bf proof of  (\ref{aMn})}

\bigskip

\fbox{\\
$\ a_1 M_1\, a_2 M_2\, \dots a_n M_n = a_1\, a_2 \dots a_n \, ({\hat R}_{12} {\hat R}_{23} \dots {\hat R}_{n-1\, n} M_n )^n\ $
\\}

\bigskip

\noindent
is based on the exchange relation
$M_1\, a_2 = a_2\, {\hat R}_1 M_2 {\hat R}_1$ (\ref{aMpm}) (here and below we denote
${\hat R}_i \equiv {\hat R}_{i\, i+1}\,$ for short) and, for $n\ge 3\,,$ on the braid relations (\ref{YBE}).
It will be made by induction. Suppose that the relation
\be
a_1 M_1 a_2 M_2 \dots a_{j-1} M_{j-1} = a_1 a_2 \dots a_{j-1} \, ({\hat R}_1 {\hat R}_2 \dots {\hat R}_{j-2} M_{j-1} )^{j-1}\ .
\lb{aMk}
\ee
holds for some $j\ge 3\,.$ It is easy to show by a direct calculation that it is valid for $j=3\,,$
\be
a_1 M_1 a_2 M_2 = a_1 ( a_2\, {\hat R}_1 M_2 {\hat R}_1 ) M_2 = a_1 a_2 ( {\hat R}_1 M_2 )^2\ ,
\lb{aM2}
\ee
and also for $j=4$ which already gives the clue to the general case:
\ba
&&a_1 M_1 \, a_2 M_2\, a_3 M_3 = a_1 ( a_2 {\hat R}_1 M_2 {\hat R}_1 ) ( a_3 {\hat R}_2 M_3 {\hat R}_2 ) M_3 =\nn\\
&&= a_1\, a_2 \, {\hat R}_1\, ( M_2 a_3 )\, {\hat R}_1 {\hat R}_2 M_3 {\hat R}_2 M_3 =\nn\\
&&= a_1\, a_2 \, {\hat R}_1\, ( a_3 {\hat R}_2 M_3 {\hat R}_2 )\, {\hat R}_1 {\hat R}_2 M_3 {\hat R}_2 M_3 = \lb{aM3}\\
&&= a_1\, a_2\, a_3\,\, {\hat R}_1 {\hat R}_2 M_3 \, ( {\hat R}_2 {\hat R}_1 {\hat R}_2 )\, M_3 {\hat R}_2 M_3 =\nn\\
&&= a_1\, a_2\, a_3\,\, {\hat R}_1 {\hat R}_2 M_3 \, ( {\hat R}_1 {\hat R}_2 {\hat R}_1 )\, M_3 {\hat R}_2 M_3 =
a_1\, a_2\, a_3\,({\hat R}_1 {\hat R}_2 M_3 )^3\ .\nn
\ea
Multiplying (\ref{aMk}) by $a_j M_j$ from the right, we first compute
\ba
&&({\hat R}_{1} \dots {\hat R}_{j-2} M_{j-1} )\, a_j =
{\hat R}_{1} \dots {\hat R}_{j-2}\, (a_j {\hat R}_{j-1}\, M_j \, {\hat R}_{j-1} ) = \nn\\
&&=a_j\,({\hat R}_{1} \dots {\hat R}_{j-2}{\hat R}_{j-1}\, M_j \, {\hat R}_{j-1} )\ ,\qquad\quad
\lb{RRMa}
\ea
which implies the relation
\be
({\hat R}_{1} \dots {\hat R}_{j-2} M_{j-1} )^{j-1}\, a_j =
a_j\,({\hat R}_{1} \dots {\hat R}_{j-1}\, M_j \, {\hat R}_{j-1} )^{j-1}\ .
\lb{RaM}
\ee
We use further the braid relations (\ref{YBE}) to derive the equality
\ba
&&{\hat R}_{j-i-1}\, ({\hat R}_{1} \dots {\hat R}_{j-1}\, M_j ) =\nn\\
&&= {\hat R}_{1} \dots {\hat R}_{j-i-1} {\hat R}_{j-i-2}{\hat R}_{j-i-1}\, {\hat R}_{j-i} \dots {\hat R}_{j-1} \, M_j  =\nn\\
&&= {\hat R}_{1} \dots {\hat R}_{j-i-2}{\hat R}_{j-i-1} {\hat R}_{j-i-2}\, {\hat R}_{j-i} \dots {\hat R}_{j-1} \, M_j  =\nn\\
&&= ( {\hat R}_{1} \dots {\hat R}_{j-1} \, M_j )\, {\hat R}_{j-i-2} \ ,\qquad i=0,1,\dots , j-3 \ .\quad
\lb{RRM}
\ea
Assuming (\ref{aMk}) and then applying (\ref{RaM}) and (\ref{RRM}), we obtain
\ba
\fl\qquad
a_1 M_1 \dots a_{j-1} M_{j-1}\, a_j M_j = a_1 \dots a_{j-1} \, ({\hat R}_1 \dots
{\hat R}_{j-2} M_{j-1} )^{j-1} a_j M_j = \nn\\
\fl\qquad
= a_1 \dots a_j \, \left( ({\hat R}_1 \dots {\hat R}_{j-1} M_j ){\hat R}_{j-1} \right)^{j-1} M_j = ... =
a_1 \dots a_j \, ({\hat R}_1 \dots {\hat R}_{j-1} M_j )^j
\lb{aMk-ind}
\ea
which proves the induction hypothesis. \eod

\bigskip

The {\bf proof of (\ref{MRn})}

\bigskip

\fbox{$\ ({\hat R}_{12} {\hat R}_{23} \dots {\hat R}_{n-1\, n} M_n )^n
= q^{1-n^2} ({\hat R}_{12} \dots {\hat R}_{n-1\, n})^n  M_{+ n}\dots M_{+ 1} M_{- 1}^{-1} \dots M_{- n}^{-1} \ $}

\bigskip

\noindent
for $M = q^{\frac{1}{n} - n} M_+ M_-^{-1}$ (\ref{M+-q}) can be made in three steps.

\smallskip

1) Define, for $j=1,\dots ,n\,,$
\ba
&&X_j := ({\hat R}_1 \dots {\hat R}_{n-1} M_{+\ n} )\, ({\hat R}_1 \dots {\hat R}_{n-2} M_{+\ n-1} {\hat R}_{n-1}) \times\dots \nn\\
&&\times({\hat R}_1 \dots {\hat R}_{n-j} M_{+ \ n-j+1} {\hat
R}_{n-j+1}\dots {\hat R}_{n-1})\, M_{-\  n-j+1}^{-1}\dots M_{-\
n}^{-1} \qquad\qquad \lb{Xk} \ea and then prove the relation \ba
&&({\hat R}_1 {\hat R}_2 \dots {\hat R}_{n-1} M_{+\ n} )\, ({\hat R}_1 {\hat R}_2 \dots {\hat R}_{n-2} M_{+\ n-1} {\hat R}_{n-1}) \times\dots \nn\\
&&\times({\hat R}_1 {\hat R}_2 \dots {\hat R}_{n-j} M_{+ \ n-j+1} {\hat R}_{n-j+1}\dots {\hat R}_{n-1}) = \nn\\
&&= ({\hat R}_1 {\hat R}_2 \dots {\hat R}_{n-1})^j M_{+\ n}\dots M_{+\ n-j+1}\ .
\lb{Xk2}
\ea
To derive (\ref{Xk2}), one has to move every $M_{+\ n-i+1}\,,\ i = 1, \dots j-1$ (starting with $M_{+\ n}\,,$ i.e. with
$i=1$) to the right until it meets the corresponding $M_{+\ n-i}\,,$ then use
$M_{+\ n-i+1} M_{+\ n-i} {\hat R}_{n-i} = {\hat R}_{n-i} M_{+\ n-i+1} M_{+\ n-i}$ (\ref{exMpm}), move further
${\hat R}_{n-i}$ to the left until it reaches the group of ${\hat R}$-s, and $M_{+\ n-i}$ to the right until
it joins the group of $M_+$-s, and repeat these steps until all ${\hat R}_1 \dots {\hat R}_{n-1}$ are brought together.

Due to (\ref{Xk2}), $X_j$ can be also written as
\be
X_j = ({\hat R}_1 {\hat R}_2 \dots {\hat R}_{n-1})^j M_{+\ n}\dots M_{+\ n-j+1} M_{-\  n-j+1}^{-1}\dots M_{-\ n}^{-1}
\lb{Xk1}
\ee
and hence, the right-hand side of (\ref{MRn}) is equal to $q^{1-n^2} X_n\,.$

\smallskip

2) Note that
\be
X_1 \equiv {\hat R}_1 \dots {\hat R}_{n-1} M_{+\ n} M_{-\ n}^{-1} =
q^{n - \frac{1}{n} }\, {\hat R}_1 \dots {\hat R}_{n-1} M_n
\lb{X1}
\ee
so that the left-hand side of (\ref{MRn}) is equal to $q^{1-n^2} X_1^n\,.$

\smallskip

3) Prove, by using
$\,M_{-\ i+1}^{-1}\, {\hat R}_i\, M_{+\ i+1}\, = \, M_{+\ i}\, {\hat R}_i\, M^{-1}_{-\ i}$ (\ref{exMpm}), that
\ba
&&M_{-\ i+1}^{-1}\, ({\hat R}_1 \dots\, {\hat R}_{i-1}\, {\hat R}_i\, M_{+\ i+1}\, {\hat R}_{i+1}\, \dots
{\hat R}_{n-1}) = \nn\\
&&= {\hat R}_1 \dots\, {\hat R}_{i-1}\,( M_{-\ i+1}^{-1}\, {\hat R}_i\, M_{+\ i+1} )\, {\hat R}_{i+1}\, \dots
{\hat R}_{n-1} = \nn\\
&&= {\hat R}_1 \dots\, {\hat R}_{i-1}\,( M_{+\ i}\, {\hat R}_i\, M^{-1}_{-\ i})\, {\hat R}_{i+1}\, \dots
{\hat R}_{n-1} = \nn\\
&&= ({\hat R}_1 \dots\, {\hat R}_{i-1}\, M_{+\ i}\, {\hat R}_i\, {\hat R}_{i+1}\, \dots
{\hat R}_{n-1})\,M_{-\ i}^{-1}
\lb{relMM}
\ea
then apply (\ref{X1}) and (\ref{relMM})\, (for $i = n-1, \dots , n-j\,$) to (\ref{Xk}) to show that
\be
X_1 X_j = X_{j+1}\ ,\quad j = 1,\dots , n-1\quad\Rightarrow\quad X_n = X_1^n\ .
\lb{Xk-rec}
\ee\eod

%%%%%%%%%%%%%%%%%%%%% Acknowledgments %%%%%%%%%%%%%%%%%%%%%%%%%

\ack{The authors thank Ivan Todorov for his interest and valuable comments on the manuscript of this work.
P.F. acknowledges the support of the Italian Ministry of University and Research (MIUR) and L.H.,
of the Bulgarian National Science Fund (grant DO 02-257).
This work has been completed during a visit of L.H. at INFN, Sezione di Trieste whose
support is gratefully acknowledged.}

%%%%%%%%%%%%%%% BIBLIOGRAPHY - (~CMP style) %%%%%%%%%%%%%%%%%%%

\end{document}